\documentclass[12pt]{article}

\usepackage{graphicx}

\usepackage[tbtags]{amsmath}
\usepackage{chicago}
\usepackage{amsfonts}
\usepackage{amssymb}

\renewcommand{\eqref}[1]{(\ref{#1})}

\newcommand{\pb}{p_{\rm b}}
\newcommand{\Uben}{U\pb{}}
\renewcommand{\sb}{s_{\rm b}}
\newcommand{\sd}{s_{\rm d}}
\newcommand{\sH}{s_{\rm H}}
\newcommand{\Ek}{{\rm E}[k]}
\newcommand{\Es}{{\rm E}[s]}

\begin{document}

\title{The Speed of Adaptation in Large Asexual Populations}
\author{Claus O. Wilke$^*$\\Digital Life Laboratory 136-93\\Caltech, Pasadena CA 91125}
\date{}

\maketitle

\noindent
Running head: Speed of adaptation
\bigskip

\bigskip

\noindent
Corresponding author:\\
Claus O. Wilke, Digital Life Laboratory 136-93, Caltech, Pasadena CA 91125\\
Phone:\ 626 395 2258, Fax: 626 395 2944, email: wilke@caltech.edu\\
\mbox{}\\
$^*$Present address:\\
Keck Graduate Institute of Applied Life Sciences, 535 Watson Drive,
Claremont, CA 91711\\
email: claus\_wilke@kgi.edu

\newpage

\noindent
\textbf{Abstract:} In large asexual populations, beneficial mutations have to
compete with each other for fixation. Here, I derive explicit analytic
expressions for the rate of substitution and the mean beneficial effect of
fixed mutations, under the assumptions that the population size $N$ is large,
that the mean effect of new beneficial mutations is smaller than the mean
effect of new deleterious mutations, and that new beneficial mutations are
exponentially distributed. As $N$ increases, the rate of substitution
approaches a constant, which is equal to the mean effect of new beneficial
mutations. The mean effect of fixed mutations continues to grow
logarithmically with $N$. The speed of adaptation, measured
as the change of log fitness over time, also grows logarithmically with $N$
for moderately large $N$, and it grows double-logarithmically for extremely
large $N$.  Moreover, I derive a simple formula that determines whether at
given $N$ beneficial mutations are expected to compete with each other or go
to fixation independently. Finally, I verify all results with
numerical simulations.

\newpage

\centerline{INTRODUCTION}
\bigskip

In asexual populations, beneficial mutations that have arisen independently in
different organisms cannot recombine and therefore have to compete for
fixation. This effect, often referred to as clonal interference
\cite{GerrishLenski98}, leads to a slowdown of adaptation for large population
sizes. A similar effect can arise in sexual populations, and is
called the Hill-Robertson effect \cite{HillRobertson66} or the traffic problem
\cite{Stephan95,KirbyStephan96}. (See also
\shortciteNP{CrowKimura65,KimuraOhta71,Barton95,Orr2000,McVeanCharlesworth2000,Gerrish2001,JohnsonBarton2002,KimStephan2003}).
Clonal interference has two main consequences: As the population size becomes
large, the increase in the rate of adaptation with increasing population size
declines, and the beneficial mutations that are fixed convey increasingly
larger beneficial effects.  A number of recent studies have tried to quantify
the rate of adaptation and the distribution of beneficial mutations in various
organisms whose predominant mode of replication is asexual, such as
\emph{Escherichia coli}
\shortcite{deVisseretal99,ImhofSchloetterer2001,Rozenetal2002}, vesicular
stomatitis virus \shortcite{Mirallesetal99,Mirallesetal2000}, and
bacteriophages $\Phi$X174 and G4
\shortcite{Bulletal2000,KichlerHolderBull2001}.

Early studies of clonal interference date back to Kimura and coworkers
\cite{CrowKimura65,KimuraOhta71}. These authors considered the same
effect $s$ for all beneficial mutations. \citeN{GerrishLenski98} were the
first to consider a distribution of beneficial effects, but neglected
deleterious mutations.  The results of \citeN{GerrishLenski98} were later
generalized by \citeN{Orr2000} to include deleterious mutations.  In the works
of both \citeN{GerrishLenski98} and \citeN{Orr2000}, the final results
(formulae for the expected substitution rate and for the mean effect of fixed
mutations) were given in the form of unwieldy double integrals, which are
difficult to interpret. [However, \citeN{GerrishLenski98} gave explicit
expressions for the unrealistic case of uniformly distributed beneficial
mutations.] From these integrals, we cannot easily estimate for what parameter
settings the interference effect becomes important, and it is unknown how the
speed of adaptation behaves for very large $N$. Moreover, even numerical
evaluation of the integrals can be tricky, because the integrand is strongly
peaked. \shortciteN{Rozenetal2002} gave an explicit expression for the
distribution of beneficial effects of fixed mutations at large $N$.  However,
this expression also does not lead to a simple expression for the mean.

Here, I derive asymptotic expansions for the expected
rate of adaptation and for the mean beneficial effect of fixed mutations,
under the assumption that beneficial mutations are distributed exponentially.
This assumption is reasonable, and has good theoretical support from
extreme-value theory \cite{Gillespie83,Gillespie91,Orr2003}. I
find that for very large $N$, the expected rate of adaptation approaches a
limiting value that is given by the mean selective advantage of new beneficial
mutations. The mean beneficial effect of fixed mutations, on the other hand,
does not reach a hard limit, but continues to grow with the logarithm of the
population size.

\bigskip

\centerline{MATERIALS AND METHODS}
\bigskip

\textbf{Model:} I consider the model analyzed by \citeN{Orr2000}.  I assume
that $N$ haploid organisms replicate asexually, and accumulate both
deleterious and advantageous mutations. The total mutation rate per genome and
generation is $U$, and the fraction of beneficial mutations is $\pb$. Hence,
the beneficial mutation rate is $\Uben$, and the deleterious mutation rate is
$U(1-\pb)$ ($\approx U$ for small $\pb$). The effects of both beneficial and
deleterious mutations are drawn from probability distributions; all mutations
act multiplicatively. I use a slightly simplified notation for beneficial and
deleterious effects of mutations in comparison to \citeN{Orr2000}: By $s$, I
denote the effect of a particular mutation, either beneficial (in which case
fitness is increased by a factor $1+s$) or deleterious (in which case fitness
is decreased by a factor $1-s$).  The mean effect of beneficial mutations is
$\sb$, and the mean effect of deleterious mutations is $\sd$. The harmonic
mean of the distribution of deleterious mutations is $\sH$. At equilibrium
(when all beneficial mutations have gone to fixation), the frequency of the
class of individuals with the highest fitness is approximately
$P_0=\exp(-U/\sH)$ \cite{Orr2000}.

I assume that beneficial mutations are exponentially distributed, that is,
beneficial effects are drawn from a distribution with probability density
function $f(s)=\exp(-s/\sb)/\sb$. The analytic calculations make no assumption
about the distribution of deleterious mutations, but all simulations have been
carried out with a truncated exponential distribution (see Simulation
methods). I assume that on average deleterious mutations have a much larger
effect than beneficial mutations ($\sb\ll\sd$), such that beneficial mutations
rarely compensate deleterious mutations.

\textbf{Simulation methods:} I carried out simulations of the model described
in the previous subsection. In the simulations, $N$ sequences were propagated
in discrete generations. The number of offspring sequences of a sequence $i$
in the next generation was binomially distributed with mean $w_i/\langle w
\rangle$, where $w_i$ is the fitness of sequence $i$ and $\langle w \rangle$
is the average fitness of the population. Each offspring sequence suffered
$k_{\rm b}$ beneficial and $k_{\rm d}$ deleterious mutations, where $k_{\rm
  b}$ and $k_{\rm d}$ were Poisson-distributed with means $U\pb$ and
$U(1-\pb)$, respectively. Each beneficial mutation increased the fitness of a
sequence by a factor of $1+s$, where $s$ was drawn from an exponential
distribution with mean $\sb$. Each deleterious mutation decreased the fitness
of a sequence by a factor of $1-s$, where $s$ was drawn from a truncated
exponential distribution with parameter $a$. The distribution was truncated
both to the left and to the right. The left truncation was necessary to avoid
a zero harmonic mean $\sH$. (For $\sH=0$, the predicted frequency of the
unmutated individuals is $P_0=0$, and the theory breaks down.) I used as a
cutoff for the left truncation the value 0.01. The right truncation is
necessary to avoid negative fitness, and here I used the cutoff value 1. As
parameter for the truncated exponential distribution, I used $a=0.1$, which
results in $\sd=0.11$ and $\sH=0.05$.

I let the population equilibrate for 1000 generations at $\pb=0$ before I set
$\pb$ to its desired value and started measuring the rate of adaptation.
Simulations were continued for up to 50,000 generations, depending on
population size (the smaller the population size, the longer the simulation
run), and replicated between 5 and 50 times (the smaller the population size,
the more replicates). For each sequence in the population, I kept track of the
number of beneficial mutations it had accumulated. At the end of a simulation
run, I subdivided the final population into classes with equal numbers of
beneficial mutations and determined the most abundant class. The number of
beneficial mutations $n$ in the most abundant class divided by the number of
generations since equilibration $\Delta t$ served as an estimator for the rate
of substitution $k$. I averaged $k$ over all replicates to obtain the result
reported for $\Ek$. In order to obtain an estimate for the change in log
fitness over time $d \log w(t)/dt$, I determined the sequence with the least
number of deleterious mutations in the most abundant class, and divided the
logarithm of the sequence's fitness by $\Delta t$. Again, I averaged over all
replicates to arrive at the values reported here. To test whether this
approach was comparable to a direct measurement of the change in population
fitness, I fitted for several exemplary runs a straight line to the logarithm
of the average population fitness as a function of time, from the end of the
equilibration time to the end of the simulation run, and took the slope of
that line as the value for $d \log w(t)/dt$. The differences in the results
obtained with these two alternative approaches were minute.  \bigskip

\centerline{RESULTS}
\bigskip

\textbf{Expected substitution rate and mean beneficial effect:}
Beneficial mutations arise in the populations at rate $N\Uben$. If this rate
is small, then they do not interfere with each other, and independently go to
fixation or are lost to drift. In this case, their expected probability of
fixation (averaged over all possible beneficial effects) is $2\sb P_0$
\cite{OrrKim98,Campos2003}, and therefore the expected rate of substitution
$\Ek$ becomes \cite{Orr2000}
\begin{equation}\label{small-N-approx}
 \Ek=2N\Uben\sb P_0\,.
\end{equation}
When beneficial mutations interfere with each other, then their probability of
fixation is reduced by a factor of $e^{-I(s)}$, with $I(s)=2 \Uben P_0 N\ln
N(s+\sb)s^{-1} e^{-s/\sb}$ \cite{GerrishLenski98,Orr2000}. $I(s)$ is the
expected number of new mutations of effect larger than $s$ that occur in the
time interval of length $t=(2/s)\ln N$ during which a mutation of effect $s$
goes to fixation. The form of $I(s)$ that I use throughout this article
assumes that beneficial mutations are distributed exponentially. The general
form for arbitrary distributions is given in \cite{GerrishLenski98,Orr2000}.
The expected rate of substitution is obtained by integrating over all
beneficial mutations. Again using the assumption that beneficial mutations are
exponentially distributed, one finds that \cite{GerrishLenski98,Orr2000}
\begin{equation}\label{Eofkexact}
\Ek=2N\Uben P_0 \sb^{-1}\int_0^\infty s e^{-I(s)-s/s_b} ds\,.
\end{equation}

\begin{figure}
\centerline{
\includegraphics[width=10cm,angle=270]{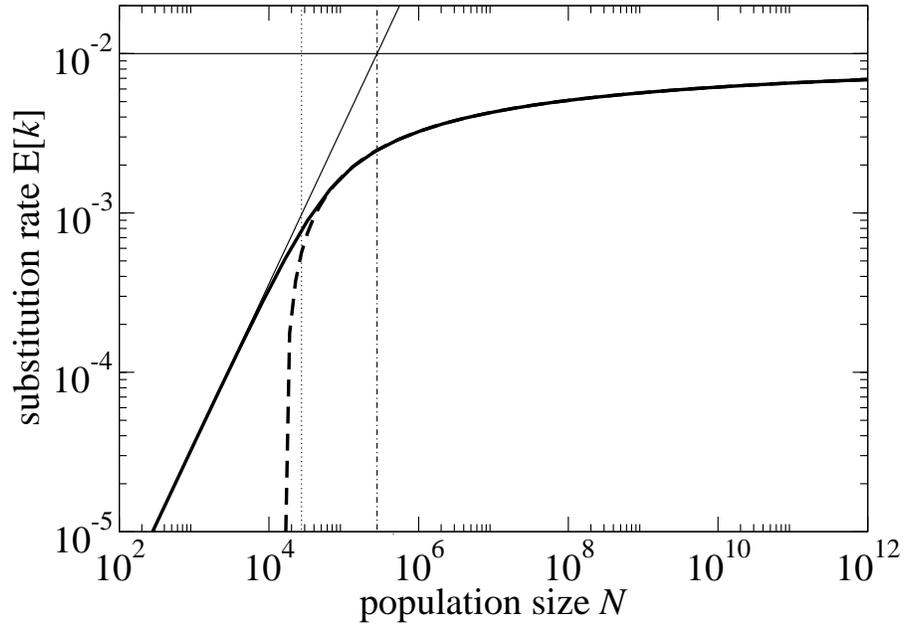}
}
\caption{Expected substitution rate $\Ek$ versus population size $N$ ($U=0.04$, $\pb=0.0001$,
  $\sb=0.01$, $\sH=0.05$). The thick solid line stems from exact
  numerical evaluation of Eq.~\eqref{Eofkexact}, and the thick dashed line
  corresponds to approximation Eq.~\eqref{large-N-approx}. The thin solid
  lines correspond to the approximations for small and large $N$,
  Eqs.~\eqref{small-N-approx} and~\eqref{very-large-N-approx}. The dash-dotted
  line indicates the onset of clonal interference according to
  Eq.~\eqref{estimate1}, and the dotted line indicates the same according to
  Eq.~\eqref{estimate2}.}\label{fig:clonal-onset}
\end{figure}

In Appendix~1, I show that for large $N$, the substitution rate becomes
\begin{equation}\label{large-N-approx}
   \Ek \approx \frac{\sb}{\ln N} [\ln(2\Uben P_0 N\ln N) +0.5772]\,.
\end{equation}
In the limit of $N\rightarrow\infty$, this expression simplifies to
\begin{equation}\label{very-large-N-approx}
  \Ek \approx \sb\,,
\end{equation}
that is, the rate of substitution reaches a hard limit that is given by the
mean beneficial effect of new mutations. Figure~\ref{fig:clonal-onset}
shows that the approximation Eq.~\eqref{large-N-approx} works well for
intermediate to large $N$. However, $\Ek$ comes close to its limiting value
$\sb$ only for very large $N$.

\begin{figure}
\centerline{
\includegraphics[width=10cm,angle=270]{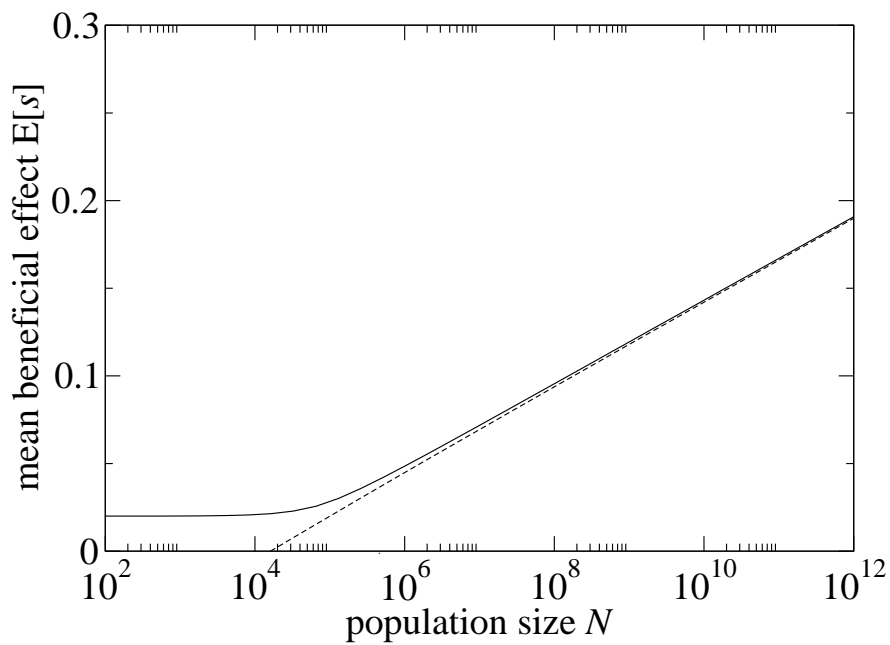}
}
\caption{Mean beneficial effect of fixed
  mutations $\Es$ versus population size $N$ ($U=0.04$, $\pb=0.0001$,
  $\sb=0.01$, $\sH=0.05$). The solid line
  stems from exact numerical evaluation of Eq.~\eqref{Eofsexact}, and the
  dashed line corresponds to approximation Eq.~\eqref{Eofs-largeN}.
  }\label{fig:exact-vs-approx}
\end{figure}

According to \citeN{GerrishLenski98}, we
can calculate the mean beneficial effect of fixed mutations $\Es$ as
\begin{equation}\label{Eofsexact}
  \Es = \frac{\int\limits_0^\infty s^2 e^{-I(s)-s/\sb}ds }{ \int\limits_0^\infty s e^{-I(s)-s/\sb}ds}\,.
\end{equation}
This expression simplifies to (see Appendix~1 for details)
\begin{equation}\label{Eofs-largeN}
   \Es \approx \sb [\ln(2 \Uben P_0 N\ln N)+0.5772]
\end{equation}
for large $N$. Figure~\ref{fig:exact-vs-approx} shows that this approximation
also works very well for intermediate to large $N$.

\textbf{Estimating the onset of clonal interference:} For small $N$, the
expected substitution rate is $\Ek\approx 2N\Uben P_0 \sb$
[Eq.~\eqref{small-N-approx}], while for very large $N$, we have $\Ek = \sb$.
On the basis of these two equations, we can derive a simple estimate for the
parameter regions in which clonal interference is relevant: We are certainly
in the clonal-interference regime if the estimate of $\Ek$ for small $N$
exceeds that for large $N$, that is, if
\begin{equation}\label{estimate1}
 N >\frac{1}{2\Uben P_0}\,.
\end{equation}
This result has a simple interpretation: Clonal interference becomes relevant
if---on average---one beneficial mutation arises in the zero-mutation class
at least every other generation. Note that the mean effect of deleterious
mutations enters this result (through $P_0$), but not the mean effect of
beneficial mutations.

The estimate Eq.~\eqref{estimate1} is fairly conservative, in the sense that
when $N$ exceeds $1/(2\Uben P_0)$, we are sure that clonal interference is
important, but clonal interference starts having some effect already for
smaller $N$. In Appendix~1, I show that an improved estimate is
\begin{equation}\label{estimate2}
 N \ln N>\frac{1}{2\Uben P_0}\,.
\end{equation}
Figure~\ref{fig:clonal-onset} illustrates where the two estimates
Eqs.~\eqref{estimate1} and \eqref{estimate2} lie with respect to the exact
expression and the approximations for $\Ek$.

\textbf{Speed of adaptation:} The expected substitution rate is in general not
an accurate measure for
 the speed of adaptation, because it disregards the
beneficial effect of the fixed mutations. A better measure is the change in
fitness (or log fitness, which is more appropriate for a multiplicative model)
over time. Clearly, the faster fitness increases, the faster a population
adapts to its environment.

As mentioned by \citeN{JohnsonBarton2002}, the change in log fitness is given
by
\begin{equation}
  \frac{d \log w(t)}{dt} = \Ek\log(1+\Es)\,.
\end{equation}
Using $\Ek=\sb$ and $\Es$ as given in Eq.~\eqref{Eofs-largeN}, we find for
large $N$
\begin{equation}\label{speed-of-adaptation-full}
  \frac{d \log w(t)}{dt} \approx \sb \ln[1+\sb\ln(2 \Uben P_0 N\ln N)+0.5772\sb]\,.
\end{equation}
This equation predicts two different regimes for $d \log w(t)/dt$, depending
on the values of $\sb$ and $N$. If $\sb$ is much smaller than one, and $N$ is
only moderately large (but sufficiently large such that $\Ek\approx\sb$), then
we can approximate $\ln(1+\Es)$ with $\Es$ and find
\begin{equation}
  \frac{d \log w(t)}{dt} \approx \sb^2[\ln(2 \Uben P_0 N\ln N)+0.5772]\,.
\end{equation}
In this regime, the speed of adaptation depends logarithmically on $N$. If on
the other hand $N$ is extremely large and $\sb$ is not extremely small, then
\begin{equation}
  \frac{d \log w(t)}{dt} \approx \sb \ln[\sb\ln(2 \Uben P_0 N\ln N)]\,.
\end{equation}
In this regime, the speed of adaptation depends double-logarithmically on $N$.

For comparison, I now calculate the speed of adaptation for small $N$. For
small $N$, clonal interference can be neglected, and therefore the mean
beneficial effect corresponds to the mean effect of beneficial mutations that
have survived drift. The distribution of these mutations is $g(s)=(s/\sb^2)
\exp(-s/\sb)$ \shortcite{Rozenetal2002,OttoJones2000}, and the mean is
$\Es=2\sb$. Using $\ln(1+\Es)\approx\Es$, we find for small $N$:
\begin{equation}
   \frac{d \log w(t)}{dt} \approx 4 \sb^2 N \Uben P_0\,.
\end{equation}
To summarize, the speed of adaptation grows linearly in $N$ for small $N$, and
logarithmically or double-logarithmically in $N$ for large $N$.
Interestingly, in the clonal interference regime, growth in the speed of
adaptation comes from the fixation of mutations with increasingly larger
effects, rather than from the fixation of increasingly more beneficial
mutations. Hence,
clonal interference slows down the speed of adaptation, but it does not lead
to a hard speed limit as long as beneficial mutations of increasingly larger
effect are accessible.

\textbf{Simulation results:} I have carried out extensive simulations to test
the accuracy of the clonal interference theory. \citeN{GerrishLenski98}
mentioned that they found good agreement between theory and simulations, but
they did not report any simulation results or the parameter regions they had
considered. \citeN{Orr2000} reported some simulation results, but his
simulations were not in the clonal interference regime [as defined by
Eq.~\eqref{estimate2}].

\begin{figure}
\centerline{
\includegraphics[width=10cm,angle=270]{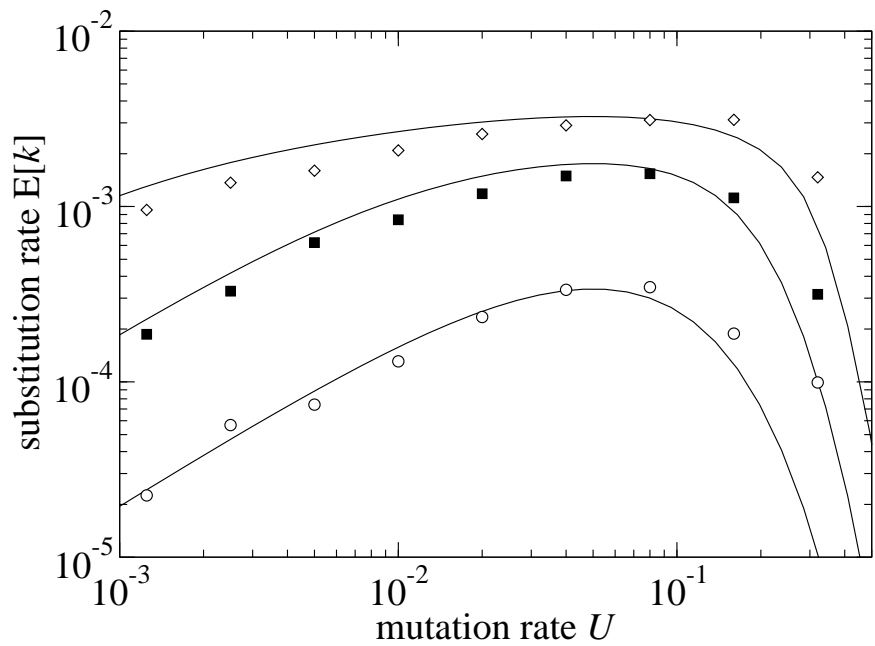}
}
\caption{Expected substitution rate $\Ek$ versus mutation rate $U$
  ($\pb=0.0001$, $\sb=0.01$, $\sH=0.05$). Population sizes are
  (from bottom to top) $N=10^4$, $N=10^5$, $N=10^6$. Solid lines
  indicate the theoretical prediction Eq.~\eqref{Eofkexact}, and points are
  simulation results.}\label{fig:substrate-vs-U}
\end{figure}

\begin{figure}
\centerline{
\includegraphics[width=10cm,angle=270]{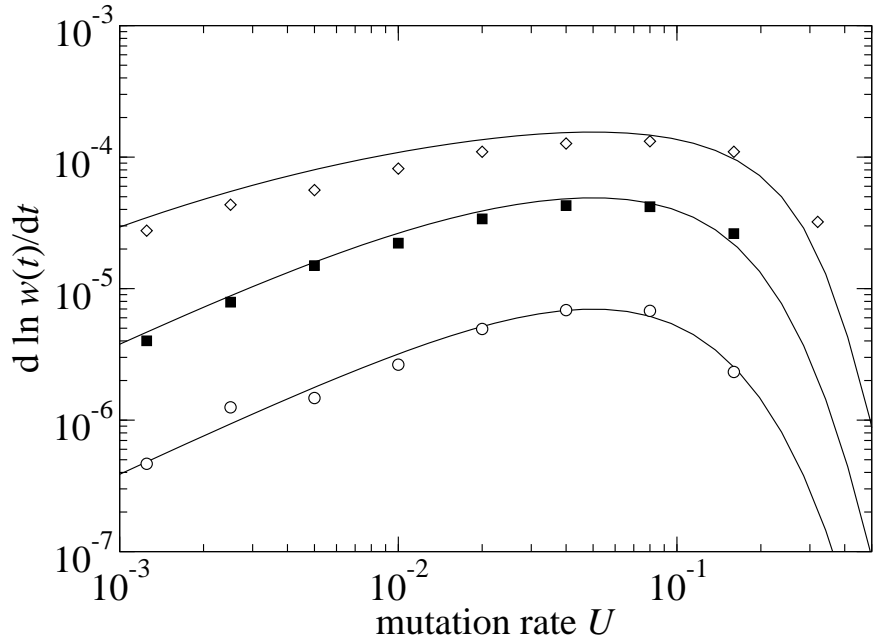}
}
\caption{Change in log fitness $d\ln w(t)/dt$ versus mutation rate $U$
  ($\pb=0.0001$, $\sb=0.01$, $\sH=0.05$). Population sizes are (from bottom to
  top) $N=10^4$, $N=10^5$, $N=10^6$. Solid lines indicate the theoretical
  prediction $\Ek\ln(1+\Es)$, and points are simulation
  results. For $U=0.4$, Muller's ratchet led to a negative $d\ln w(t)/dt$ in
  the populations of size $N=10^4$ and $N=10^5$. The corresponding two data
  points are therefore missing from this figure.}\label{fig:dwdt-vs-U}
\end{figure}

Figure~\ref{fig:substrate-vs-U} shows the expected substitution rate as a
function of the mutation rate $U$, both as predicted by Eq.~\eqref{Eofkexact}
and as found in simulations. Below the optimal mutation rate $U=\sH$ at which
the substitution rate assumes its maximum \cite{Orr2000}, agreement between
theory and data is good over a wide range of population sizes. Above
$U=\sH$, the theory underestimates $\Ek$. This effect is caused by the
accumulation of slightly deleterious mutations in the simulations. The theory
assumes that only those beneficial mutations that arise in backgrounds free
from deleterious mutations can go to fixation. However, for large $U$,
sequences that carry one or several slightly deleterious mutations become so
frequent that it becomes likely that one of them acquires a beneficial
mutation of sufficiently large effect to compensate the deleterious
background, and goes to fixation. The degree to which the theory
underestimates $\Ek$ increases as $\sH$ decreases. In the limit of $\sH=0$,
the theory predicts that $\Ek=0$, while simulations show that the true results
(with identical $\sb$ and $\sd$) are not substantially different from those
shown in Fig.~\ref{fig:substrate-vs-U} (data not shown).  Surprisingly, the
theory accurately predicts the change in log fitness $d\ln w(t)/dt$ even in
the regime of large $U$, as long as $d\ln w(t)/dt$ is not negative
(Fig.~\ref{fig:dwdt-vs-U}). [For very high $U$, Muller's ratchet
\cite{Muller64,Felsenstein74,Haigh78,GordoCharlesworth2000} becomes the
predominant force in the dynamic of the evolving population, and the change in
log fitness can assume negative values.] Apparently, in the regime of large
$U$, the theory underestimates $\Ek$ and overestimates $\Es$, in such a way
that the two effects nearly cancel each other.

\begin{figure}
\centerline{
\includegraphics[width=10cm,angle=270]{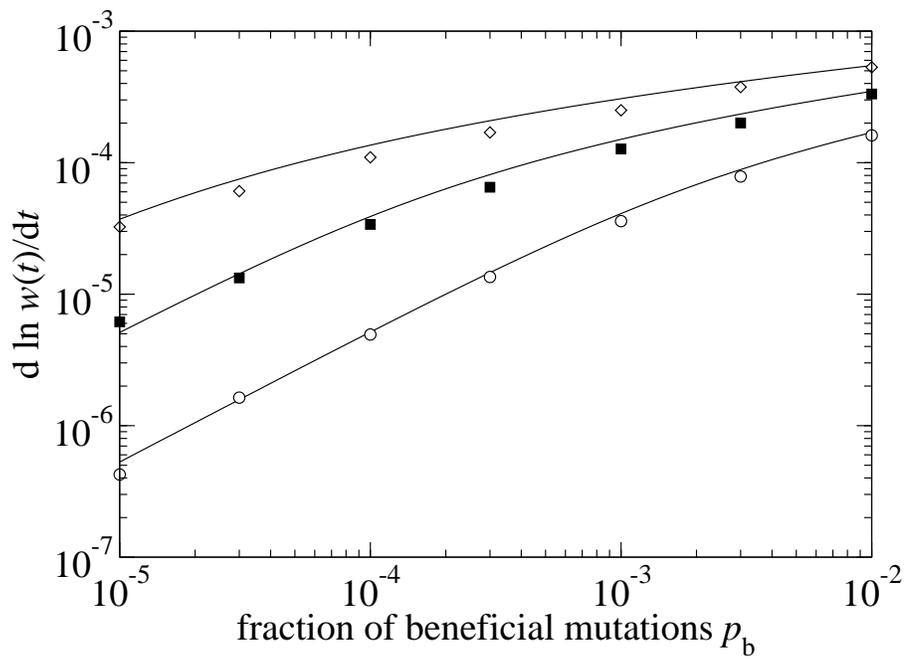}
}
\caption{Change in log fitness $d\ln w(t)/dt$ versus fraction of beneficial
  mutations $\pb$ ($U=0.02$, $\sb=0.01$, $\sH=0.05$). Population
  sizes are (from bottom to top) $N=10^4$, $N=10^5$, $N=10^6$. Solid lines
  indicate the theoretical prediction $\Ek\ln(1+\Es)$, and points are
  simulation results.}\label{fig:dwdt-vs-pb}
\end{figure}

In Fig.~\ref{fig:dwdt-vs-pb}, I show the change in log fitness as a function
of the fraction of beneficial mutations $\pb$. Again, we see excellent
agreement between theory and simulation. However, for $\pb\gtrsim0.001$ the
theory underestimates $\Ek$ and overestimates $\Es$, in such a way that the
two effects cancel each other (data not shown).

\begin{figure}
\centerline{
\includegraphics[width=10cm,angle=270]{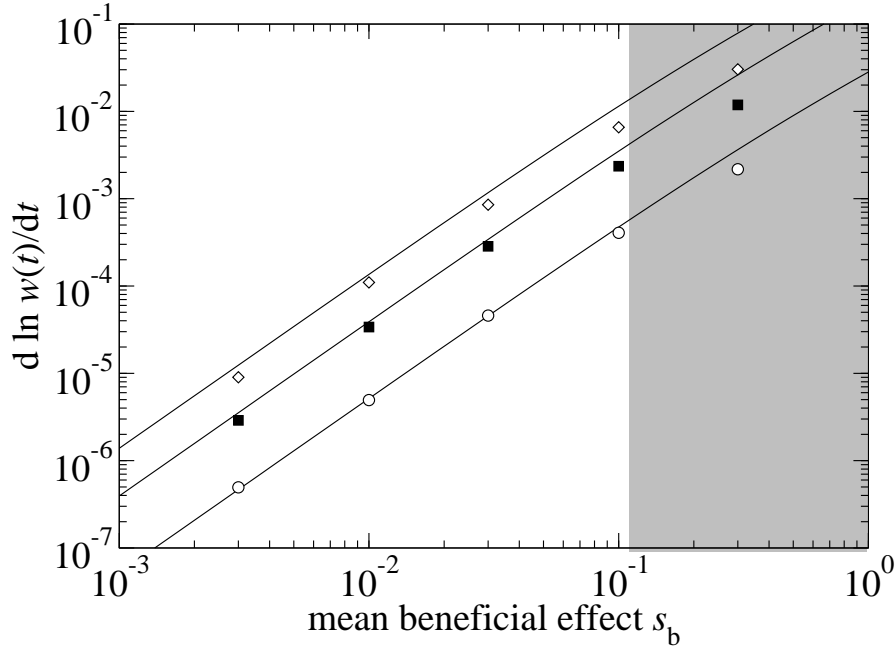}
}
\caption{Change in log fitness $d\ln w(t)/dt$ versus mean effect of
  new beneficial mutations $\sb$ ($\pb=0.0001$, $U=0.02$, $\sH=0.05$).
  Population sizes are (from bottom to top) $N=10^4$, $N=10^5$, $N=10^6$.
  Solid lines indicate the theoretical prediction $\Ek\ln(1+\Es)$, and points
  are simulation results. In the shaded region, the mean effect of new
  beneficial mutations $\sb$ exceeds the mean effect of new deleterious
  mutations $\sd$.}\label{fig:dwdt-vs-sb}
\end{figure}

Finally, in Fig.~\ref{fig:dwdt-vs-sb}, I show the change in log fitness as a
function of the mean effect of new beneficial mutations $\sb$, while holding
the mean effect of new deleterious mutations $\sd$ constant at $\sd=0.11$
($\sH=0.05$). As shown in Appendix~1, the theory predicts that both $\Ek$
and $\Es$ should depend linearly on $\sb$ for all parameter values. The change
in log fitness should therefore depend quadratically on $\sb$ for $\sb\lesssim
1$, which means that $d\ln w(t)/dt$ should appear approximately as a straight
line with slope 2 in the double-logarithmic plot. We see that the simulation
data agree very well with the theory as long as $\sb\lesssim\sd$, but start to
diverge slowly as $\sb$ grows larger than $\sd$.

\bigskip

\centerline{DISCUSSION}
\bigskip

Clonal interference is often said to impose a speed limit on adaptation. Here,
I have shown that the speed of adaptation, measured as the change in log
fitness over time, does not reach a hard limit, but continues to grow even for
very large $N$. This growth is fueled by the discovery of mutations with ever
larger beneficial effect in large populations, rather than by an increase in
the rate of substitutions.

My results hinge on the assumption that new beneficial mutations are
exponentially distributed. If beneficial mutations are distributed such that
large effects are absent, then the rate of adaptation will most likely reach
an upper limit for large $N$. If on the other hand beneficial effects follow a
distribution with long tail (such as a power-law or Cauchy distribution), then
the speed of adaptation may grow even faster than predicted by
Eq.~\eqref{speed-of-adaptation-full} for large $N$. To date, we do not have a
good understanding of the true distribution of beneficial effects in
experimental systems. However, an exponential distribution has good
theoretical support \cite{Gillespie83,Gillespie91,Orr2003}, has led to good
agreement between theory and experiment in \emph{E.~coli}
\shortcite{Rozenetal2002}, and overall seems to be a reasonable choice.

Arguments for an exponential distribution of new deleterious mutations are
not as strong. At the same time, the theory is much less dependent on the
particulars of the distribution of deleterious mutations. As long as we have
an accurate expression for $P_0$, and beneficial mutations are unlikely to
compensate deleterious mutations, the theory should work. In practice, this
means that the theory should work with any distribution that does not produce
an excessive amount of slightly deleterious mutations. (Neutral mutations
could be dealt with by considering them as a reduction in the overall mutation
rate $U$.)

De Visser et al.\ (1999) measured the speed of adaptation in \emph{E.~coli},
varying both the population size and the mutation rate each over approximately
two orders of magnitude. They found that the speed of adaptation did not grow
in proportion to increases in population size or mutation rate. In fact, apart
from the experiments carried out at the lowest mutation rate, the speed of
adaptation changed only very little with population size or mutation rate.
These results indicate that the populations with the larger size and higher
mutation rates could not benefit from the additional beneficial mutations that
must have appeared.  The results of \shortciteN{deVisseretal99} thus provide
good support for the clonal interference model on a qualitative level.
Quantitatively, however, their data seem to disagree with the model analyzed
here: Fig.~2A of \shortciteN{deVisseretal99} suggests that the speed of
adaptation runs quickly into a hard limit, whereas the model predicts that the
speed should continue to grow logarithmically, at least with respect to
population size.

There are two reasons that may have caused this discrepancy. First,
\shortciteN{deVisseretal99} plotted the speed of adaptation versus the
relative mutation-supply rate (which is the product of population size and
relative mutation rate). Such a plot is problematic, because the
mutation-supply rate does not uniquely specify the speed of adaptation in the
clonal interference model. [In order for the mutation-supply rate to uniquely
specify the speed of adaptation, population size and mutation rate would have
to enter the equations always as a product, which is not the case. In
Eq.~\eqref{speed-of-adaptation-full}, for example, the term $P_0$ depends on
$U$ but not on $N$, while the term $\ln N$ does not depend on $U$.] The model
predicts that the speed of adaptation should increase with increasing $N$,
whereas it should reach a maximum and then decrease with increasing $U$.
Thus, a plot of the speed of adaptation versus population size (at fixed $U$)
is inherently more informative than a plot of the speed of adaptation versus
mutation rate (at fixed $N$). In the latter case, a decline in the
increase of the speed of adaptation may also indicate that the mutation rate
approaches the optimal mutation rate $U=\sH$. Since
\shortciteN{deVisseretal99} studied only two different population sizes (and
three mutation rates), it is not possible to replot a subset of their data
versus population size at fixed $U$ and obtain a quantitative comparison to
the model.

Second, the speed of adaptation at a high mutation-supply rate may have been
reduced (thus giving the impression of a hard speed limit) in part because the
populations began to run out of beneficial mutations. De Visser et al.\
propagated the populations for 1000 generations, and determined the speed of
adaptation from the total fitness increase over these 1000 generations. In
particular for the large population size, it seems that adaptation slowed down
considerably after 500 generations (\shortciteNP{deVisseretal99}, Fig.~1B).
[Note, however, that this argument does not invalidate the overall conclusion
of \shortciteN{deVisseretal99}. The fitness increase after 200 generations in
their Fig.~1 shows strong dependence on the mutation rate for the small
population size, and weak dependence on the mutation rate for the large
population size, which agrees very well with the predictions of the clonal
interference model.] A related explanation for the apparent slowdown in
the speed of adaptation at large population size is that the large populations
may have found mutations of large beneficial effect earlier in the experiments
than the small populations, as predicted by the clonal interference model.

The clonal interference model assumes an infinite supply of beneficial
mutations, and this assumption is of course unrealistic. Nevertheless, we can
expect good agreement between model and experiment if the experiment is
restricted to a relatively short number of generations, or if only the effect
of the first fixed mutation is measured. An experiment of the latter kind was
carried out by \shortciteN{Rozenetal2002}, who found that the measured
distribution of beneficial effects in \emph{E.~coli} was in good agreement
with the distribution as predicted by the clonal interference model.

The data of \shortciteN{Rozenetal2002} also allows us to estimate the onset of
clonal interference in \emph{E.~coli}. By fitting the theoretical prediction
for the distribution of beneficial effects to their data,
\shortciteN{Rozenetal2002} derived estimates for the mean beneficial effect of
new mutations $\sb$ and for the beneficial mutation rate $\Uben$. They found
$\sb=0.024$ and $\Uben=5.9\times 10^{-8}$. Having an estimate for the
beneficial mutation rate, we can use Eq.~\eqref{estimate2} to estimate the
population size in these \emph{E.~coli} populations at which clonal
interference becomes important. Since we do not have a good estimate for the
mutational load in these populations, we set $P_0=1$, which means that we
neglect the effect of deleterious mutations. [See \citeN{Orr2000} for a
discussion of this problem and its implications for the estimates of $\sb$ and
$\Uben$.] As a consequence, we most likely underestimate the population size
at which clonal interference becomes important. Further, instead of the factor
2 in front of $\Uben$, we use 0.6. This factor takes into account that the
\emph{E.~coli} populations fluctuate in size under standard laboratory
conditions, see \shortciteN{Rozenetal2002}, p.~1044. Thus, we use for our
estimate $N \ln N>1/(0.6\times5.9\times 10^{-8})$. This condition simplifies
to $N>2\times 10^{6}$.  [Using a less accurate method based on only the
expected substitution rate and mean beneficial effect of fixed mutations,
\citeN{GerrishLenski98} had earlier derived an estimate of $\Uben=2\times
10^{-9}$, which leads to $N>4.7\times 10^7$.]  \shortciteN{Rozenetal2002}
carried out their experiments at an effective population size of $N=3.3\times
10^7$, which means that clonal interference probably had an effect on their
results. This reasoning is consistent with the observation that the mean
beneficial effect of fixed mutations is clearly larger than $2\sb$ in their
data (\shortciteNP{Rozenetal2002}, Fig. 3).

Clonal interference has not only been studied in \emph{E. coli}, but also in
vesicular stomatitis virus (VSV). Following \shortciteN{deVisseretal99},
\shortciteN{Mirallesetal99} fitted a linear and a hyperbolic model to the rate
of fitness change in VSV as a function of population size, and found that the
hyperbolic model provided the better fit.  However, their data does not
plateau at high $N$, and visual inspection of their Fig.~1 suggests that a
logarithmic model might fit their data as well.  On the other hand, the
multiplicity of infection was changed alongside with the population size in
these experiments, so that a slow-down in the speed of adaptation could also
be due to increased virus-virus interactions within cells, rather than clonal
interference \cite{WilkeNovella2003}.

My simulations have shown that the theory of clonal interference works well
for small to moderate mutation rates, but fails at high mutation
rates, when Muller's ratchet becomes important. Another effect at high
mutation rates that is neglected in the theory (but was also absent from the
simulations) is the evolution of mutational robustness: If the distribution of
deleterious mutations itself can change, then at a high mutation rate there is
a selective pressure to minimize the mutational load of the population
\shortcite{vanNimwegenetal99b,Wilkeetal2001b,WilkeAdami2003}. This effect will
increase the mean fitness of the population, and will also increase the
potential for further adaptation by increasing $P_0$. In general, at a high
mutation rate we have to consider that a mutation will be combined with
additional mutations on the way to fixation. Therefore, we cannot simply
assume that a mutation of beneficial effect $s$ has probability of fixation
$2s$, but have to use fairly complicated mathematical tools (such as
multi-type branching processes) to calculate the fixation probability
\shortcite{Barton95,JohnsonBarton2002,Wilke2003,Iwasaetal2004}. As a
consequence, it is unlikely that we will ever have a simple closed-form
expression for the speed of adaptation at a high mutation rate.

A second regime in which the theory---not surprisingly---breaks down is when
$\sb$ exceeds $\sd$. In this regime, we cannot simply neglect all beneficial
mutations that do not arise on genetic backgrounds free of deleterious
mutations. \citeN{JohnsonBarton2002} have recently studied this situation, but
not in the clonal interference regime. In the clonal interference regime,
there are two opposing effects to be considered: On the one hand, the total
number of competing mutations should increase, since now beneficial mutations
on deleterious backgrounds compete for fixation as well. On the other hand,
the total fitness effect of many of the mutations that are competing for
fixation is smaller than what we expect from the distribution of new
beneficial mutations, because the beneficial effects are reduced by
deleterious backgrounds. Without a detailed analysis, it is unclear which of
the two effects is more important. However, if the results from the present
study are an indication, in the clonal interference regime the number of
competing mutations will be less important than the distribution of their
beneficial effects, which means that the present theory should overestimate
the rate of adaptation for $\sb>\sd$. Indeed, I observed exactly this behavior
in my simulations (Fig.~\ref{fig:dwdt-vs-sb}).

Whether the assumption $\sb<\sd$ is reasonable is not yet resolved. (Likely,
the answer to this question will also depend on the particular species under
study and on the concrete selection regime.) Several authors found that
deleterious mutations were frequent but had a very small effect
\shortcite{Mukaietal72,Ohnishi77,KibotaLynch96,Shabalinaetal97,ElenaMoya99},
while others found that deleterious mutations were less frequent but of larger
effect
\shortcite{Keightley96,FernandezLopezFanjul96,GarciaDorado97,KeightleyCaballero97}.
If the first set of results is representative, then beneficial mutations may
indeed have on average a larger effect on fitness than deleterious mutations.
While it is reasonably straightforward to study the distribution of
deleterious mutations in mutation-accumulation experiments, it is much harder
to measure the distribution of new beneficial mutations (as opposed to the
distribution of fixed beneficial mutations, which is skewed towards mutations
of large effect).  However, evidence from experimental evolution with viruses
shows that in some cases, beneficial mutations must have very large effects:
\shortciteN{Wichmanetal99} found several-thousand-fold increase in population
growth after ten days of selection in phage $\Phi$X174, and
\shortciteN{Novellaetal95} found fitness increases by a factor of 10 or more
within five to ten generations in vesicular stomatitis virus. In both cases,
the observed fitness increase within a very short time frame can only be
explained by a large supply of beneficial mutations of large effect.

To summarize, in this contribution I have found the following novel
conclusions:
\begin{enumerate}
\item The expected rate of adaptation approaches the mean beneficial effect of
  new mutations for large $N$.
\item The mean beneficial effect of fixed mutations grows logarithmically in
  $N$ for large $N$.
\item Clonal interference effects become important if $N\ln N$ is larger than
  $1/(2\Uben P_0)$.
\item The speed of adaptation grows logarithmically in $N$ for moderately
  large $N$, and double-logarithmically for extremely large $N$.
\item For large $N$, the speed of adaptation is limited by the distribution of
  beneficial effects of new mutations rather than by the supply rate of new
  mutations.
\end{enumerate}

\bigskip

\centerline{ACKNOWLEDGMENTS}
\bigskip

I would like to thank H.A.~Orr for clarifications regarding his work, and
S.F.~Elena, R.E.~Lenski, S.~P. Otto, J.A.G.M. de~Visser, and two anonymous
reviewers for helpful comments on earlier versions of this article.  This
work was supported by the NSF under contract No.~DEB-9981397.


\newpage

\begin{appendix}
\centerline{APPENDIX 1}
\bigskip

\textbf{Expected substitution rate and mean beneficial effect:} First, we
notice that $\Ek$ depends linearly on the mean effect of beneficial mutations:
Substituting $x$ for $s/\sb$ in Eq.~\eqref{Eofkexact}, and writing $A=2 \Uben
P_0 N\ln N$, we find
\begin{equation}\label{Eofkexact-no-sb}
  \Ek = \frac{\sb A }{\ln N} \int\limits_0^\infty x \exp[-A (1+1/x) e^{-x}
  -x]dx \,.
\end{equation}
Therefore, the shape of $\Ek$ as a function of $N$ or $U$ is independent of
the value of $\sb$.

We are interested in an asymptotic expansion of the integral in
Eq.~\eqref{Eofkexact-no-sb} for large $N$. For the asymptotic expansion to
work, $N$ needs to be so large that $A$ is large. Clearly, for any given
$\Uben P_0$, we can always choose $N$ sufficiently large such that $A$ is
large.  Because of the exponential factor in the integrand, the main
contribution to the integral comes from values of $x$ for which
$A(1+1/x)e^{-x}+x$ is small. Since the first term decays exponentially with
$x$ while the second term grows linearly, in general the main contribution to
the integral will come from small $x$. However, for large $A$, the first term
becomes small only when $x$ is substantially larger than one. In this regime,
we can neglect the term $1/x$, and the integral in Eq.~\eqref{Eofkexact-no-sb}
is then identical to the integral $J_n(A)$ defined in Appendix~2
with $n=1$. Using the expression for $J_1(A)$ given in
Eq.~\eqref{eq:J2-result}, we find
\begin{equation}
   \Ek \approx \frac{\sb}{\ln N} [\ln(2\Uben P_0 N\ln N) +\gamma]\,.
\end{equation}
In the limit of very large $N$, we obtain the even simpler expression
$\Ek \approx \sb$.

Using a reasoning similar to that for the expected substitution rate, we find
that the expected beneficial effect [as given in Eq.~\eqref{Eofsexact}] also
depends linearly on $\sb$, and simplifies for large $N$ to $\Es=\sb
J_2(A)/J_1(A)$ (again with $A=2 \Uben P_0 N\ln N$). Using the expressions
given in Eqs.~\eqref{eq:J1-result} and~\eqref{eq:J2-result}, we find
\begin{equation}
  \Es \approx \sb [\ln(2 \Uben P_0 N\ln N)+\gamma]+\frac{\sb\pi^2/6}{\ln(2 \Uben P_0 N\ln N)+\gamma}\,.
\end{equation}
In the limit of very large $N$, the second term disappears, and we end up with
\begin{equation}
   \Es \approx \sb [\ln(2 \Uben P_0 N\ln N)+\gamma]\,.
\end{equation}

\textbf{Estimating the onset of clonal interference:} We can derive an
estimate of the parameter region in which clonal interference becomes
important by calculating the point at which the approximation for $\Ek$ for
small $N$ [Eq.~\eqref{small-N-approx}] comes the closest to the approximation
for $\Ek$ for large $N$, Eq.~\eqref{large-N-approx}. Since the shape of $\Ek$
is not influenced by $\sb$ (see above), we can set $\sb=1$ for this
calculation. Further, we write $C=2 \Uben P_0$. Now, we have to find the
minimum of the function
\begin{equation}
  g(C, N) = C N - [\ln(C N \ln N)+\gamma]/\ln N\,.
\end{equation}
We find $\partial g(C,N)/\partial C = N-1/(C \ln N)$, which leads to the
condition
\begin{equation}
 N \ln N>\frac{1}{C}
\end{equation}
for the onset of clonal interference. (Differentiating with respect to $N$
yields approximately the same condition). This condition cannot be solved for
$N$ in a closed-form expression, but is easy to evaluate numerically.
\bigskip

\centerline{APPENDIX 2}
\bigskip

\textbf{Integrals:} For the asymptotic expansion, we have to solve integrals
of the form
\begin{equation}\label{eq:Jn-def}
  J_n(A) = \int\limits_0^\infty x^n \exp(-Ae^{-x} -x) dx\,,
\end{equation}
in particular for the cases $n=1$ and $n=2$. After substituting
$z=Ae^{-x}$, we obtain
\begin{align}
  J_n(A) &= \frac{1}{A}\int\limits_0^A (\ln A - \ln z)^k e^{-z}
  dz\notag\\
  &= \frac{1}{A} \sum_{k=0}^n \binom{n}{k} (\ln A)^{n-k}(-1)^k\int\limits_0^A (\ln z)^k e^{-z}
  dz\,.
\end{align}
The main contribution to the remaining integral comes from small $z$, while
$A$ is large in the cases considered here. Therefore, we can replace the upper
limit of integration with $\infty$. For the three relevant cases $k=0$, $k=1$,
and $k=2$, the integrals are $\int_0^\infty e^{-z} dz = 1$, $\int_0^\infty \ln
z\, e^{-z} dz = -\gamma$, $\int_0^\infty (\ln z)^2 e^{-z} dz = \gamma^2
+\pi^2/6$, where $\gamma\approx0.5772$ is the Euler constant. Thus, we
find approximately
\begin{align}\label{eq:J1-result}
  J_1(A) &= (\ln A + \gamma)/A \,,\\\label{eq:J2-result}
  J_2(A) &= [(\ln A + \gamma)^2 +\pi^2/6]/A\,.
\end{align}
\end{appendix}

\cleardoublepage
\newpage

\end{document}